\documentclass[%
 preprint,
superscriptaddress,
 amsmath,amssymb,
 aps,
pre,
]{revtex4-1}
\usepackage{braket}
\usepackage{color}
\usepackage{graphicx}
\usepackage{dcolumn}
\usepackage{bm}
\usepackage{subfig}
\usepackage{algorithmic}
\begin{document}

\preprint{}

\title{Generalized Fourier transform method for solving nonlinear  anomalous diffusion equations}

\author{Jie Yao}
\altaffiliation[Previously at  ]{University of Houston, Department of Mechanical Engineering}
\email[Email:]{jie.yao@ttu.edu}
\affiliation{
Texas Tech University, Department of Mechanical Engineering, Lubbock, Texas, USA, 79409}

\author{Cameron L. Williams}
\affiliation{University of Houston, Department of Mathematics, Houston, Texas, USA,77204}

\author{Fazle Hussain}
\affiliation{
Texas Tech University, Department of Mechanical Engineering, Lubbock, Texas, USA, 79409}

\author{Donald J. Kouri}
\email[Email:]{kouri@central.uh.edu}
\affiliation{University of Houston, Department of  Mathematics, Houston, Texas, USA,77204}
\affiliation{University of Houston, Departments of Mechanical Engineering and
Physics, Houston, Texas, USA,77204}

\date{\today}

\begin{abstract}
The solution of a nonlinear  diffusion equation is numerically investigated using the generalized Fourier transform method.
This equation includes fractal dimensions and power-law dependence on the radial variable and on the diffusion function. 
The generalized Fourier transform  approach is the extension  of the Fourier transform method used 
for the normal diffusion equation.
The feasibility of the  approach is validated  by 
comparing the numerical result with the exact solution for 
a point-source. 
The merit of the numerical method is that it provides a way to 
calculate anomalous diffusion with an arbitrary initial condition.


\end{abstract}

\maketitle


\section{\label{sec:level1} Introduction}
In the last few decades, anomalous diffusion has been extensively studied 
in a variety of physical applications, 
such as turbulent diffusion \cite{gavrilov1995trial},  surface growth  \cite{spohn1993surface}, 
transport of fluid in porous media \cite{spohn1993surface}, hydraulics problems \cite{daly2004similarity},
etc.   
The diffusion is usually characterized by the time dependence of mean-square displacement (MSD) viz.,
$\braket{r^2}\propto t^\sigma$. 
The MSD grows linearly with time ($\sigma=1$) for the normal diffusion case.
The process is called sub-diffusion for $0<\sigma<1$ and  super-diffusion for $\sigma>1$.
The standard normal diffusion described by the Gaussian distribution 
can be obtained from the usual Fokker-Planck equation with a constant diffusion coefficient and zero drift \cite{risken1984fokker}. 
Extensions of the conventional Fokker-Planck equation have been used to study anomalous diffusion.
For example, anomalous diffusion can be obtained by the usual Fokker-Planck equation, but with a 
variable diffusion coefficient \cite{fa2005exact,fa2011solution}. 
It can also be achieved by incorporating nonlinear terms in the diffusion term, or external forces 
\cite{bologna2000anomalous,assis2006nonlinear,lenzi2010some, zola2008exact}. 
In some approaches,  fractional equations have been employed to 
analyze anomalous diffusion and related phenomena 
\cite{metzler2000random,sokolov2002fractional,tsallis2002anomalous,liu2004numerical}.

In this paper, we study the generalized nonlinear diffusion equation including a fractal dimension $d$ 
and a diffusion coefficient 
which depends on the radial variable and the diffusion function  $\rho$
\cite{malacarne2001nonlinear,pedron2002nonlinear,abraham2005lie} 
\begin{eqnarray} \label{eqn:gd}
\frac{\partial \rho}{\partial t}=K_0 \frac{1}{r^{d-1}}\frac{\partial}{\partial r}\Big(r^{d-1-\theta}\frac{\partial}{\partial r}\rho^\nu\Big),
\end{eqnarray}
with the initial and boundary conditions  
\begin{eqnarray}
 \rho(r,t_0)&=&\rho_0(r),\\
 \rho(\infty,t)&=&0,
\end{eqnarray}
where $r$ is the radial coordinate, 
and $\theta$ and $\nu$ are real parameters.
When the  diffusion coefficient  is a function of $r$ only, it is a generalization of the diffusion equation for fractal geometry \cite{o1985analytical}. 
It is the  traditional nonlinear diffusion equation when the diffusion coefficient  depends on $\rho$ only \cite{crank1979mathematics,bluman1980remarkable}. 
Analytical solutions of eq. (\ref{eqn:gd}) with a point source
have been reported in \cite{malacarne2001nonlinear,pedron2002nonlinear}, 
where an ansatz for $\rho$ is proposed as a  general stretched Gaussian function.
In \cite{abraham2005lie}, the same analytic solutions were also obtained by using Lie group symmetry analysis. 

Motivated by the research on  generalized nonlinear diffusion, we propose here a numerical 
method for solving eq. (\ref{eqn:gd}) using a generalized Fourier transform.
The generalized Fourier transform (also called the $\Phi_n$ transform) is  a new family of integral transforms developed by 
Willams et. al. \cite{williams2017fourier,williams2017}. These transforms share all the properties of the Fourier transform; hence can be employed to perform  more general frequency and time-frequency analysis \cite{kouri2018point,8270495}.

In section \ref{sec:2}, a brief introduction of  the generalized Fourier transform is provided. 
The procedure of using the generalized Fourier transform for solving the generalized nonlinear diffusion equation 
is discussed in \ref{sec:level2} and \ref{sec:level3}.
The method is validated by comparison between analytical and numerical results. 
Then, some numerical results for a non-Delta function initial condition are given in \ref{sec:level4}. Conclusions are drawn in \ref{sec:level5}.

\section{generalized Fourier transform}\label{sec:2}

The generalized Fourier transform $\Phi_n$ is defined as 
\begin{eqnarray}
 \Phi_nf(k)=\int\varphi_n(k x)f(x) \mathrm{d}x,
\end{eqnarray}
where the integral kernel $\varphi_n(\omega x)=c_n(k x)+\mathrm{i} s_n(k x)$ is,
\begin{eqnarray}
 c_n(\eta)=\frac{1}{2}|\eta|^{n-1/2}J_{-1+\frac{1}{2n}}\Big(\frac{|\eta|^n}{n}\Big),
\end{eqnarray}
and 
\begin{eqnarray}
 s_n(\eta)=\frac{1}{2}\mathrm{sgn}(\eta)|\eta|^{n-1/2}J_{1-\frac{1}{2n}}\Big(\frac{|\eta|^n}{n}\Big),
\end{eqnarray}
where $J_\nu(\eta)$ is the cylindrical Bessel function, and $n$ is the transform order, i.e., $n=1,2,\cdots$. 

The Fourier transform $\mathcal{F}$ is the special case with $n=1$. 
The $\Phi_n$ transform shares many properties of the Fourier transform. 
Here we focus on two properties which will be used later.
It is well-known that the Fourier transform preserves the functional form of a Gaussian; particularly, 
$\mathcal{F}[g_1]=g_1$ if $g_1(x)=\exp(-x^2/2)$. For the generalized Fourier transform, we have
$\Phi_ng_n=g_n$, if $g_n(x)=e^{-\frac{x^{2n}}{2n}}$.

In addition, the generalized Fourier transform also has the following derivative property:
\begin{eqnarray}\label{eqn:derivative}
 \Phi_n[\frac{\partial}{\partial x} x^{2-2n} \frac{\partial }{\partial x } f]=k^{-2n} \Phi_nf.
\end{eqnarray}

In  \cite{williams2014fourier}, the $\Phi_n$ transform is developed for integer order case. 
However, it can  be easily  extended  to the  non-integer case;  
see \cite{williams2014fourier} for additional discussion  of the properties of the $\Phi_n$ transform. 

\section{solving the generalized diffusion equation with generalized Fourier transform}
It is well known that the Fourier transform can be used to find the solution for the standard diffusion equation \cite{haberman1983elementary}. 
Motivated by this idea, here we explore   
employing the generalized Fourier transform for solving the generalized nonlinear diffusion equation. 

\subsection{The O'Shaugnessy-Procaccia
anomalous diffusion equation on fractals}\label{sec:level2}
Let us first  consider the  generalization of the diffusion equation for fractal geometry, where the diffusion coefficient is a function of $r$ only (i.e.,  $\nu=1$) \cite{o1985analytical}. Eq. (\ref {eqn:gd}) can be reduced to
\begin{eqnarray}\label{eqn:rgd}
\frac{\partial \rho (r,t)}{\partial t}= \frac{K_0}{r^{d-1}}
\Big(\frac{\partial}{\partial r}r^{d-1-\theta}\frac{\partial}{\partial r}\rho (r,t)\Big).
\end{eqnarray}
In order to perform the $\Phi_n$ transform, we apply the following scaling relationship
\begin{eqnarray}
 \frac{\partial}{\partial r}(\cdot)=dr^{d-1}\frac{\partial}{\partial r^d}(\cdot),
\end{eqnarray}
to eq. (\ref{eqn:rgd}); and with some simplification, we obtain 
\begin{eqnarray}
\frac{\partial \rho (\tilde{r},t)}{\partial t}=\tilde{K}_0
\frac{\partial}{\partial \tilde{r}}\tilde{r}^{2-\lambda /d}\frac{\partial}{\partial \tilde{r}}\rho (\tilde{r},t),
\end{eqnarray}
where $\tilde{K}_0=K_0d^2$, $\tilde{r}=r^d$, and $\lambda=2+\theta$.

By applying the  $\Phi_n$ transform to both sides and employing the derivative identity (eq. (\ref{eqn:derivative})), 
we obtain the diffusion equation in the wavenumber domain
\begin{eqnarray}\label{eqn:rgdk}
\frac{\partial \tilde{\rho} }{\partial t}=- \tilde{K}_0k^{\lambda /d}\tilde{\rho},
\end{eqnarray}
with $\tilde{\rho}= \Phi_n\rho$.

Eq. (\ref{eqn:rgdk}) can be  exactly solved as
\begin{eqnarray} \label{eqn:ana}
 \tilde{\rho}(k,t)=e^{-\tilde{K}_0k^{\lambda /d}t} \tilde{\rho}_0.
\end{eqnarray}
The solution to eq. (\ref{eqn:rgd}) is then obtained by applying the inverse $\Phi_n$ transform 
to $\tilde{\rho}(k,t)$.

We validate the $\Phi_n$ transform method by comparing the numerical results with the analytical solution or a point source at the origin (i.e., $\rho(r,t_0)=\delta(r)$), which is given as \cite{o1985analytical}
\begin{eqnarray}
 \rho_a (r,t)=\frac{\lambda}{d\varGamma(d/\lambda))}\Big(\frac{1}{K_0\lambda^2t}\Big)^{d/\lambda}
 \exp\Big(-\frac{r^\lambda}{K_0\lambda^2t}\Big).\nonumber\\
\end{eqnarray}

\begin{figure}
        \centering
	 \includegraphics[width=0.45\textwidth]{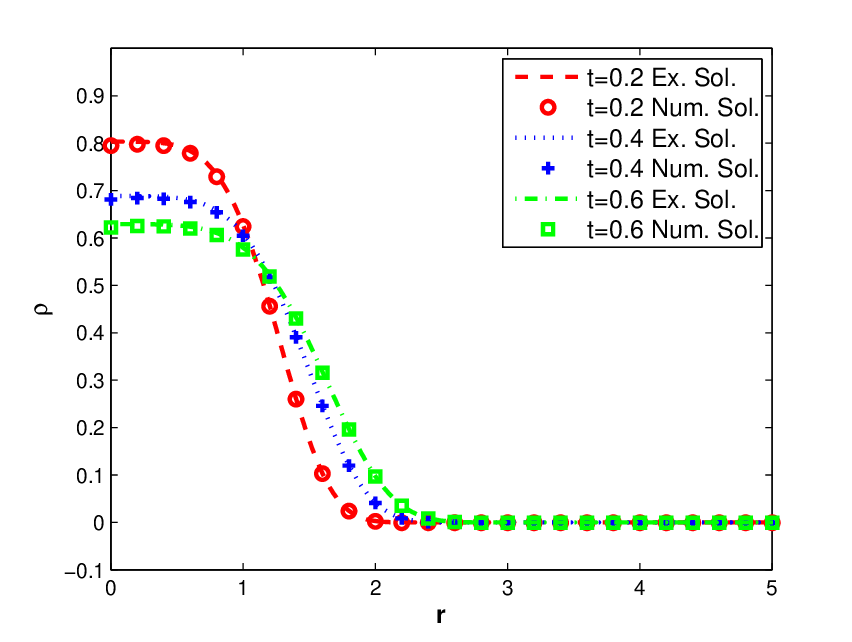}	
	  \caption{Comparison between exact (line) and numerical (symbols) solutions  with 
	  $K_0=1$, $D=1$ and $\theta=2.5$.}
	   \label{fig:sm1} 
\end{figure}

Fig. (\ref{fig:sm1}) shows the analytical  and the numerical solution 
for $K_0=1$, $D=1$, and $\theta=2.5$ at different times.
According to the classification discussed in \cite{malacarne2001nonlinear}, 
this example is a subdiffusion case with $\theta>D(1-\nu)$.
From fig. (\ref{fig:sm1}), it can be seen that the numerical solution 
is in good agreement with the analytical solution. 
In addition, we observe the  short tail behaviours of  the solution $ \rho (r,t)$.

\subsection{Generalized nonlinear equation}\label{sec:level3}
Now we consider the generalized nonlinear diffusion equation with $\nu\neq1$.
For the point source ( or Dirac delta initial condition), 
Eq. (\ref{eqn:gd}) was analytically 
solved using a generalized stretched Gaussian function approach in \cite{malacarne2001nonlinear}:
\begin{eqnarray}
 \rho(r,t)=\begin{cases} 
 [1-(1-q)\beta(t)r^\lambda]^{1/(1-q)}/Z(t),\\ 
  \qquad  \qquad  \qquad \text{if }1-(1-q)\beta(t)r^\lambda \geq0;\nonumber\\
 0,  \qquad  \qquad \quad\text{otherwise}.
 \end{cases}
\end{eqnarray}
Here $q=2-\nu$, and $\beta(t)$ and $Z(t)$ are functions  given in eq. (12) in \cite{malacarne2001nonlinear}. 
The same solution is derived in \cite{pedron2002nonlinear} using Lie group symmetry method. 

In order to  solve the generalized nonlinear diffusion equation numerically,
we follow the procedure in \ref{sec:level2},  transforming the spatial domain 
equation to the wavenumber domain using the $\Phi_n$ transform. 
Instead of Eq. (\ref{eqn:rgdk}), the wavenumber domain diffusion equation becomes 
\begin{eqnarray}\label{eqn:rgdk2}
\frac{\partial \widetilde{\rho} }{\partial t}=- \widetilde{K}_0k^{\lambda/d}\widetilde{\rho^\nu},
\end{eqnarray}
with $\widetilde{\rho^\nu}= \Phi_n(\rho^\nu)$.

\begin{figure}
\centering
 \subfloat[]{
	     \label{fig:sm3:a}
	  \includegraphics[width=0.45\textwidth]{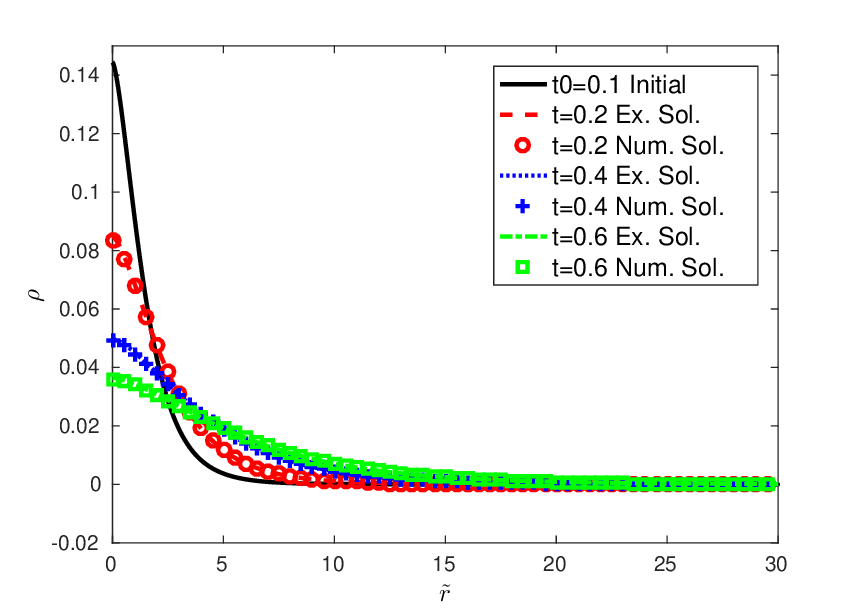}}\\
	   \subfloat[]{
	      \label{fig:sm3:b}
	  \includegraphics[width=0.45\textwidth]{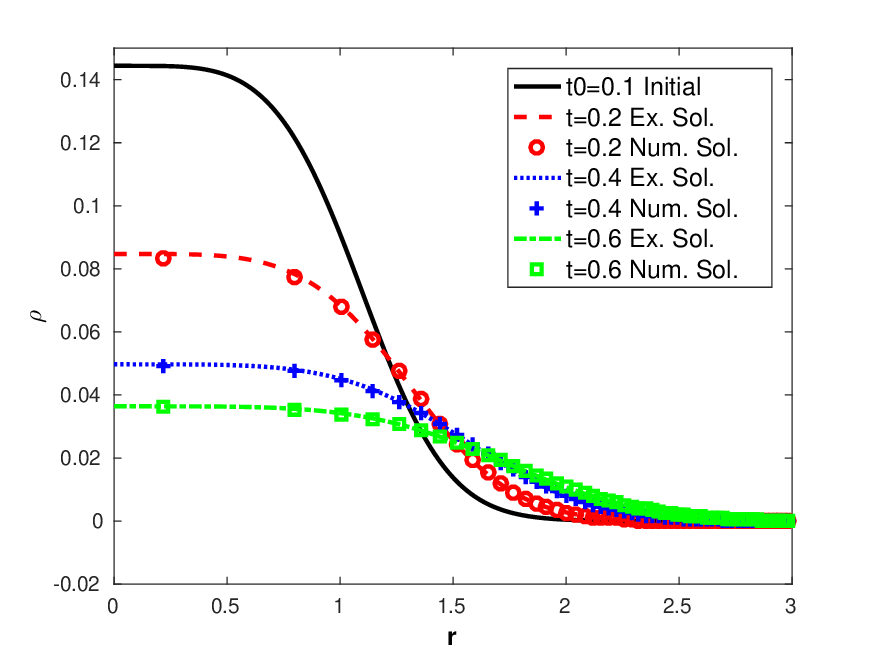}}\\

\caption{Comparison between exact (line) and numerical (symbols) solutions  in (a) scaled coordinate
and (b)  original coordinate with$K_0=1$, $D=3$, $\theta=2.5$ and $\nu=0.8$.}
\label{fig:sm3}
\end{figure}

Due to the presence of nonlinearity term in the right hand side of Eq. \ref{eqn:rgdk2} (i.e., $\rho^\nu$),  
an analytical solution in the form of Eq. \ref{eqn:ana} is difficult to be obtained. 
However, Eq. \ref{eqn:rgdk2} can be numerically solved by employing certain types  of 
time-stepping discretization methods for the time derivative. Here, the simple forward Euler finite difference scheme is employed for time discretization with $\Delta t=0.01$s. 
Equally spaced mesh with $N_r=1001$ is used for  the domain size $\tilde{r}=[0,30]$. 
The comparisons between the exact \cite{malacarne2001nonlinear,pedron2002nonlinear}
and numerical solution  for the point source (dirac Delta function $\delta(r)$)  initial condition in scaled ($\tilde{r}$) and original ($r$) coordinates
are shown in figs. (\ref{fig:sm3:a}) and (\ref{fig:sm3:b}), respectively. The parameters used here  are $K_0=1$, $D=3$, $\theta=2.5$, and $\nu=0.8$.
Note that to avoid performing a $\Phi_n$ transform for the fractional order of Dirac Delta function (as the definition of $\delta^\nu$ is also an ongoing research topic \cite{li2016powers,jarad2018defining}), 
we use $\rho_a (r,t_0=0.1)$ as the initial condition for our numerical simulation. Again, good agreement between the numerical and analytical solutions can been observed. 

\subsection{Generalized diffusion for arbitrary initial condition}\label{sec:level4}
\begin{figure}
        \centering
	 \includegraphics[width=0.45\textwidth]{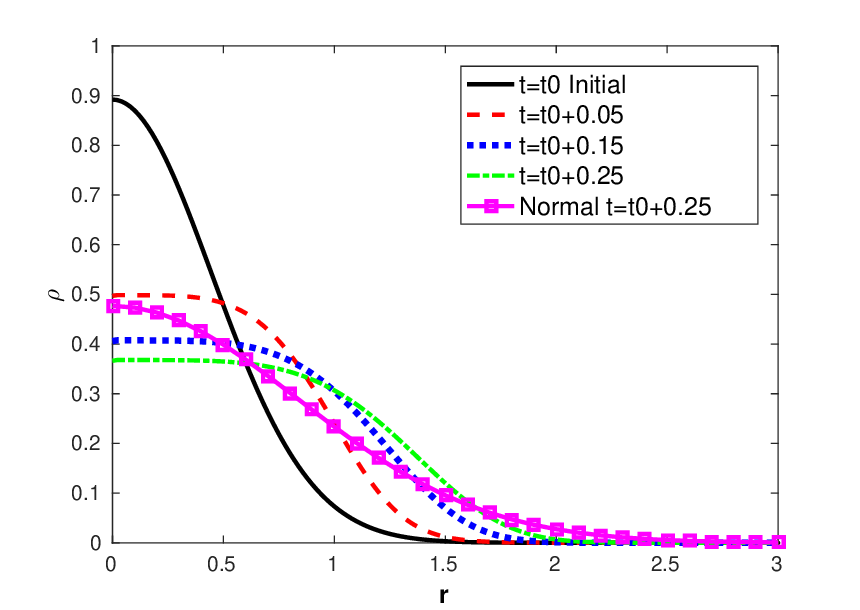}	
	  \caption{Numerical solution with 
	  $K_0=1$, $D=1$ and $\theta=2.5$ for the initial condition eq. (\ref{eqn:ninitial}).  The solution for normal diffusion with the same initial condition ($\square$ symbol) is also included for comparison.}
	   \label{fig:sm5} 
\end{figure}

The merit of the numerical approach using the generalized Fourier transform is that it  provides a way for solving the generalized diffusion equation with arbitrary initial condition. 
In fig. (\ref{fig:sm5}), we present the numerical solution of the generalized diffusion equation
for   $K_0=1$, $D=1$ and $\theta=2.5$
with the  Gaussian initial condition

\begin{eqnarray} \label{eqn:ninitial}
 \rho_0 (r,t_0)=\frac{1}{\sqrt{4\pi t_0}} \exp(-\frac{r^2}{4K_0 t_0}),
\end{eqnarray}
where $t_0=0.1$. 

As we can see, the diffusion process finally approaches 
the same generalized Gaussian shape as in the point source case (fig. (\ref{fig:sm1})).
In \cite{anteneodo2006long}, it  was analytically shown that
the normal diffusion equation, when initialized with a generalized Gaussian distribution will
asymptotically approach its final solution, i.e., a Gaussian distribution. 
Here, we present  a numerical example of  what amounts to the ``generalized central limit'' behaviour 
in which the diffusion process will finally transform the arbitrary initial distribution
 to the corresponding generalized Gaussian distribution \cite{toscani2005central,schwammle2008q}. 
 A rigorous proof of the existence of the acctractor of the generalized Gaussian diffusion has been done \cite{kouri2017point}.
However, as mentioned in \cite{schwammle2008q}, the diffusion procedure, initialized with different distribution,
may take very long time to reach its asymptotic behaviour. 
In addition, by comparing with the solution for normal diffusion with the same initial condition,
the sub-diffusion process clearly exhibits the short tail behaviour. 

\section{Conclusion}
\label{sec:level5}
In this paper, a numerical method for solving the generalized nonlinear diffusion equation has been
presented and validated.
The method is based on the generalized Fourier transform $\Phi_n$
and  has been validated by comparing the numerical solution with  analytical solution for the point source.
The presented method may serve as a useful tool to study a variety of systems 
involving  the anomalous diffusion.
Currently, no fast transform algorithm has yet been developed for the  $\Phi_n$ transform.
This issue will be investigated in a future study.

\begin{acknowledgments}
Discussions with Bernhard G. Bodmann are appreciated. Partial support for this work was provided by resources of the uHPC cluster
managed by the University of Houston under NFS Award Number 1531814.
\end{acknowledgments}



\bibliographystyle{apsrev4-1}
\bibliography{apssamp}

\begin{thebibliography}{32}%
\makeatletter
\providecommand \@ifxundefined [1]{%
 \@ifx{#1\undefined}
}%
\providecommand \@ifnum [1]{%
 \ifnum #1\expandafter \@firstoftwo
 \else \expandafter \@secondoftwo
 \fi
}%
\providecommand \@ifx [1]{%
 \ifx #1\expandafter \@firstoftwo
 \else \expandafter \@secondoftwo
 \fi
}%
\providecommand \natexlab [1]{#1}%
\providecommand \enquote  [1]{``#1''}%
\providecommand \bibnamefont  [1]{#1}%
\providecommand \bibfnamefont [1]{#1}%
\providecommand \citenamefont [1]{#1}%
\providecommand \href@noop [0]{\@secondoftwo}%
\providecommand \href [0]{\begingroup \@sanitize@url \@href}%
\providecommand \@href[1]{\@@startlink{#1}\@@href}%
\providecommand \@@href[1]{\endgroup#1\@@endlink}%
\providecommand \@sanitize@url [0]{\catcode `\\12\catcode `\$12\catcode
  `\&12\catcode `\#12\catcode `\^12\catcode `\_12\catcode `\%12\relax}%
\providecommand \@@startlink[1]{}%
\providecommand \@@endlink[0]{}%
\providecommand \url  [0]{\begingroup\@sanitize@url \@url }%
\providecommand \@url [1]{\endgroup\@href {#1}{\urlprefix }}%
\providecommand \urlprefix  [0]{URL }%
\providecommand \Eprint [0]{\href }%
\providecommand \doibase [0]{http://dx.doi.org/}%
\providecommand \selectlanguage [0]{\@gobble}%
\providecommand \bibinfo  [0]{\@secondoftwo}%
\providecommand \bibfield  [0]{\@secondoftwo}%
\providecommand \translation [1]{[#1]}%
\providecommand \BibitemOpen [0]{}%
\providecommand \bibitemStop [0]{}%
\providecommand \bibitemNoStop [0]{.\EOS\space}%
\providecommand \EOS [0]{\spacefactor3000\relax}%
\providecommand \BibitemShut  [1]{\csname bibitem#1\endcsname}%
\let\auto@bib@innerbib\@empty
\bibitem [{\citenamefont {Gavrilov}\ \emph {et~al.}(1995)\citenamefont
  {Gavrilov}, \citenamefont {Klepikova},\ and\ \citenamefont
  {Rodean}}]{gavrilov1995trial}%
  \BibitemOpen
  \bibfield  {author} {\bibinfo {author} {\bibfnamefont {V.}~\bibnamefont
  {Gavrilov}}, \bibinfo {author} {\bibfnamefont {N.}~\bibnamefont {Klepikova}},
  \ and\ \bibinfo {author} {\bibfnamefont {H.}~\bibnamefont {Rodean}},\
  }\href@noop {} {\bibfield  {journal} {\bibinfo  {journal} {Atmospheric
  Environment}\ }\textbf {\bibinfo {volume} {29}},\ \bibinfo {pages} {2317}
  (\bibinfo {year} {1995})}\BibitemShut {NoStop}%
\bibitem [{\citenamefont {Spohn}(1993)}]{spohn1993surface}%
  \BibitemOpen
  \bibfield  {author} {\bibinfo {author} {\bibfnamefont {H.}~\bibnamefont
  {Spohn}},\ }\href@noop {} {\bibfield  {journal} {\bibinfo  {journal} {Journal
  de Physique I}\ }\textbf {\bibinfo {volume} {3}},\ \bibinfo {pages} {69}
  (\bibinfo {year} {1993})}\BibitemShut {NoStop}%
\bibitem [{\citenamefont {Daly}\ and\ \citenamefont
  {Porporato}(2004)}]{daly2004similarity}%
  \BibitemOpen
  \bibfield  {author} {\bibinfo {author} {\bibfnamefont {E.}~\bibnamefont
  {Daly}}\ and\ \bibinfo {author} {\bibfnamefont {A.}~\bibnamefont
  {Porporato}},\ }\href@noop {} {\bibfield  {journal} {\bibinfo  {journal}
  {Physical Review E}\ }\textbf {\bibinfo {volume} {70}},\ \bibinfo {pages}
  {056303} (\bibinfo {year} {2004})}\BibitemShut {NoStop}%
\bibitem [{\citenamefont {Risken}(1984)}]{risken1984fokker}%
  \BibitemOpen
  \bibfield  {author} {\bibinfo {author} {\bibfnamefont {H.}~\bibnamefont
  {Risken}},\ }in\ \href@noop {} {\emph {\bibinfo {booktitle} {The
  Fokker-Planck Equation}}}\ (\bibinfo  {publisher} {Springer},\ \bibinfo
  {year} {1984})\ pp.\ \bibinfo {pages} {63--95}\BibitemShut {NoStop}%
\bibitem [{\citenamefont {Fa}(2005)}]{fa2005exact}%
  \BibitemOpen
  \bibfield  {author} {\bibinfo {author} {\bibfnamefont {K.~S.}\ \bibnamefont
  {Fa}},\ }\href@noop {} {\bibfield  {journal} {\bibinfo  {journal} {Physical
  Review E}\ }\textbf {\bibinfo {volume} {72}},\ \bibinfo {pages} {020101}
  (\bibinfo {year} {2005})}\BibitemShut {NoStop}%
\bibitem [{\citenamefont {Fa}(2011)}]{fa2011solution}%
  \BibitemOpen
  \bibfield  {author} {\bibinfo {author} {\bibfnamefont {K.~S.}\ \bibnamefont
  {Fa}},\ }\href@noop {} {\bibfield  {journal} {\bibinfo  {journal} {Physical
  Review E}\ }\textbf {\bibinfo {volume} {84}},\ \bibinfo {pages} {012102}
  (\bibinfo {year} {2011})}\BibitemShut {NoStop}%
\bibitem [{\citenamefont {Bologna}\ \emph {et~al.}(2000)\citenamefont
  {Bologna}, \citenamefont {Tsallis},\ and\ \citenamefont
  {Grigolini}}]{bologna2000anomalous}%
  \BibitemOpen
  \bibfield  {author} {\bibinfo {author} {\bibfnamefont {M.}~\bibnamefont
  {Bologna}}, \bibinfo {author} {\bibfnamefont {C.}~\bibnamefont {Tsallis}}, \
  and\ \bibinfo {author} {\bibfnamefont {P.}~\bibnamefont {Grigolini}},\
  }\href@noop {} {\bibfield  {journal} {\bibinfo  {journal} {Physical Review
  E}\ }\textbf {\bibinfo {volume} {62}},\ \bibinfo {pages} {2213} (\bibinfo
  {year} {2000})}\BibitemShut {NoStop}%
\bibitem [{\citenamefont {Assis~Jr}\ \emph {et~al.}(2006)\citenamefont
  {Assis~Jr}, \citenamefont {da~Silva}, \citenamefont {da~Silva}, \citenamefont
  {Lenzi},\ and\ \citenamefont {Lenzi}}]{assis2006nonlinear}%
  \BibitemOpen
  \bibfield  {author} {\bibinfo {author} {\bibfnamefont {P.}~\bibnamefont
  {Assis~Jr}}, \bibinfo {author} {\bibfnamefont {P.}~\bibnamefont {da~Silva}},
  \bibinfo {author} {\bibfnamefont {L.}~\bibnamefont {da~Silva}}, \bibinfo
  {author} {\bibfnamefont {E.}~\bibnamefont {Lenzi}}, \ and\ \bibinfo {author}
  {\bibfnamefont {M.}~\bibnamefont {Lenzi}},\ }\href@noop {} {\bibfield
  {journal} {\bibinfo  {journal} {Journal of mathematical physics}\ }\textbf
  {\bibinfo {volume} {47}},\ \bibinfo {pages} {3302} (\bibinfo {year}
  {2006})}\BibitemShut {NoStop}%
\bibitem [{\citenamefont {Lenzi}\ \emph {et~al.}(2010)\citenamefont {Lenzi},
  \citenamefont {Lenzi}, \citenamefont {Gimenez},\ and\ \citenamefont
  {da~Silva}}]{lenzi2010some}%
  \BibitemOpen
  \bibfield  {author} {\bibinfo {author} {\bibfnamefont {E.}~\bibnamefont
  {Lenzi}}, \bibinfo {author} {\bibfnamefont {M.}~\bibnamefont {Lenzi}},
  \bibinfo {author} {\bibfnamefont {T.}~\bibnamefont {Gimenez}}, \ and\
  \bibinfo {author} {\bibfnamefont {L.}~\bibnamefont {da~Silva}},\ }\href@noop
  {} {\bibfield  {journal} {\bibinfo  {journal} {Journal of Engineering
  Mathematics}\ }\textbf {\bibinfo {volume} {67}},\ \bibinfo {pages} {233}
  (\bibinfo {year} {2010})}\BibitemShut {NoStop}%
\bibitem [{\citenamefont {Zola}\ \emph {et~al.}(2008)\citenamefont {Zola},
  \citenamefont {Lenzi}, \citenamefont {Evangelista}, \citenamefont {Lenzi},
  \citenamefont {Lucena},\ and\ \citenamefont {da~Silva}}]{zola2008exact}%
  \BibitemOpen
  \bibfield  {author} {\bibinfo {author} {\bibfnamefont {R.}~\bibnamefont
  {Zola}}, \bibinfo {author} {\bibfnamefont {M.}~\bibnamefont {Lenzi}},
  \bibinfo {author} {\bibfnamefont {L.}~\bibnamefont {Evangelista}}, \bibinfo
  {author} {\bibfnamefont {E.}~\bibnamefont {Lenzi}}, \bibinfo {author}
  {\bibfnamefont {L.}~\bibnamefont {Lucena}}, \ and\ \bibinfo {author}
  {\bibfnamefont {L.}~\bibnamefont {da~Silva}},\ }\href@noop {} {\bibfield
  {journal} {\bibinfo  {journal} {Physics Letters A}\ }\textbf {\bibinfo
  {volume} {372}},\ \bibinfo {pages} {2359} (\bibinfo {year}
  {2008})}\BibitemShut {NoStop}%
\bibitem [{\citenamefont {Metzler}\ and\ \citenamefont
  {Klafter}(2000)}]{metzler2000random}%
  \BibitemOpen
  \bibfield  {author} {\bibinfo {author} {\bibfnamefont {R.}~\bibnamefont
  {Metzler}}\ and\ \bibinfo {author} {\bibfnamefont {J.}~\bibnamefont
  {Klafter}},\ }\href@noop {} {\bibfield  {journal} {\bibinfo  {journal}
  {Physics reports}\ }\textbf {\bibinfo {volume} {339}},\ \bibinfo {pages} {1}
  (\bibinfo {year} {2000})}\BibitemShut {NoStop}%
\bibitem [{\citenamefont {Sokolov}\ \emph {et~al.}(2002)\citenamefont
  {Sokolov}, \citenamefont {Klafter},\ and\ \citenamefont
  {Blumen}}]{sokolov2002fractional}%
  \BibitemOpen
  \bibfield  {author} {\bibinfo {author} {\bibfnamefont {I.~M.}\ \bibnamefont
  {Sokolov}}, \bibinfo {author} {\bibfnamefont {J.}~\bibnamefont {Klafter}}, \
  and\ \bibinfo {author} {\bibfnamefont {A.}~\bibnamefont {Blumen}},\
  }\href@noop {} {\bibfield  {journal} {\bibinfo  {journal} {Physics Today}\
  }\textbf {\bibinfo {volume} {55}},\ \bibinfo {pages} {48} (\bibinfo {year}
  {2002})}\BibitemShut {NoStop}%
\bibitem [{\citenamefont {Tsallis}\ and\ \citenamefont
  {Lenzi}(2002)}]{tsallis2002anomalous}%
  \BibitemOpen
  \bibfield  {author} {\bibinfo {author} {\bibfnamefont {C.}~\bibnamefont
  {Tsallis}}\ and\ \bibinfo {author} {\bibfnamefont {E.}~\bibnamefont
  {Lenzi}},\ }\href@noop {} {\bibfield  {journal} {\bibinfo  {journal}
  {Chemical Physics}\ }\textbf {\bibinfo {volume} {284}},\ \bibinfo {pages}
  {341} (\bibinfo {year} {2002})}\BibitemShut {NoStop}%
\bibitem [{\citenamefont {Liu}\ \emph {et~al.}(2004)\citenamefont {Liu},
  \citenamefont {Anh},\ and\ \citenamefont {Turner}}]{liu2004numerical}%
  \BibitemOpen
  \bibfield  {author} {\bibinfo {author} {\bibfnamefont {F.}~\bibnamefont
  {Liu}}, \bibinfo {author} {\bibfnamefont {V.}~\bibnamefont {Anh}}, \ and\
  \bibinfo {author} {\bibfnamefont {I.}~\bibnamefont {Turner}},\ }\href@noop {}
  {\bibfield  {journal} {\bibinfo  {journal} {Journal of Computational and
  Applied Mathematics}\ }\textbf {\bibinfo {volume} {166}},\ \bibinfo {pages}
  {209} (\bibinfo {year} {2004})}\BibitemShut {NoStop}%
\bibitem [{\citenamefont {Malacarne}\ \emph {et~al.}(2001)\citenamefont
  {Malacarne}, \citenamefont {Mendes}, \citenamefont {Pedron},\ and\
  \citenamefont {Lenzi}}]{malacarne2001nonlinear}%
  \BibitemOpen
  \bibfield  {author} {\bibinfo {author} {\bibfnamefont {L.}~\bibnamefont
  {Malacarne}}, \bibinfo {author} {\bibfnamefont {R.}~\bibnamefont {Mendes}},
  \bibinfo {author} {\bibfnamefont {I.}~\bibnamefont {Pedron}}, \ and\ \bibinfo
  {author} {\bibfnamefont {E.}~\bibnamefont {Lenzi}},\ }\href@noop {}
  {\bibfield  {journal} {\bibinfo  {journal} {Physical Review E}\ }\textbf
  {\bibinfo {volume} {63}},\ \bibinfo {pages} {030101} (\bibinfo {year}
  {2001})}\BibitemShut {NoStop}%
\bibitem [{\citenamefont {Pedron}\ \emph {et~al.}(2002)\citenamefont {Pedron},
  \citenamefont {Mendes}, \citenamefont {Malacarne},\ and\ \citenamefont
  {Lenzi}}]{pedron2002nonlinear}%
  \BibitemOpen
  \bibfield  {author} {\bibinfo {author} {\bibfnamefont {I.~T.}\ \bibnamefont
  {Pedron}}, \bibinfo {author} {\bibfnamefont {R.}~\bibnamefont {Mendes}},
  \bibinfo {author} {\bibfnamefont {L.~C.}\ \bibnamefont {Malacarne}}, \ and\
  \bibinfo {author} {\bibfnamefont {E.~K.}\ \bibnamefont {Lenzi}},\ }\href@noop
  {} {\bibfield  {journal} {\bibinfo  {journal} {Physical Review E}\ }\textbf
  {\bibinfo {volume} {65}},\ \bibinfo {pages} {041108} (\bibinfo {year}
  {2002})}\BibitemShut {NoStop}%
\bibitem [{\citenamefont {Abraham-Shrauner}(2005)}]{abraham2005lie}%
  \BibitemOpen
  \bibfield  {author} {\bibinfo {author} {\bibfnamefont {B.}~\bibnamefont
  {Abraham-Shrauner}},\ }\href@noop {} {\bibfield  {journal} {\bibinfo
  {journal} {Journal of Physics A: Mathematical and General}\ }\textbf
  {\bibinfo {volume} {38}},\ \bibinfo {pages} {2547} (\bibinfo {year}
  {2005})}\BibitemShut {NoStop}%
\bibitem [{\citenamefont {O'Shaughnessy}\ and\ \citenamefont
  {Procaccia}(1985)}]{o1985analytical}%
  \BibitemOpen
  \bibfield  {author} {\bibinfo {author} {\bibfnamefont {B.}~\bibnamefont
  {O'Shaughnessy}}\ and\ \bibinfo {author} {\bibfnamefont {I.}~\bibnamefont
  {Procaccia}},\ }\href@noop {} {\bibfield  {journal} {\bibinfo  {journal}
  {Physical review letters}\ }\textbf {\bibinfo {volume} {54}},\ \bibinfo
  {pages} {455} (\bibinfo {year} {1985})}\BibitemShut {NoStop}%
\bibitem [{\citenamefont {Crank}(1979)}]{crank1979mathematics}%
  \BibitemOpen
  \bibfield  {author} {\bibinfo {author} {\bibfnamefont {J.}~\bibnamefont
  {Crank}},\ }\href@noop {} {\emph {\bibinfo {title} {The mathematics of
  diffusion}}}\ (\bibinfo  {publisher} {Oxford university press},\ \bibinfo
  {year} {1979})\BibitemShut {NoStop}%
\bibitem [{\citenamefont {Bluman}\ and\ \citenamefont
  {Kumei}(1980)}]{bluman1980remarkable}%
  \BibitemOpen
  \bibfield  {author} {\bibinfo {author} {\bibfnamefont {G.}~\bibnamefont
  {Bluman}}\ and\ \bibinfo {author} {\bibfnamefont {S.}~\bibnamefont {Kumei}},\
  }\href@noop {} {\bibfield  {journal} {\bibinfo  {journal} {Journal of
  Mathematical Physics}\ }\textbf {\bibinfo {volume} {21}},\ \bibinfo {pages}
  {1019} (\bibinfo {year} {1980})}\BibitemShut {NoStop}%
\bibitem [{\citenamefont {Williams}\ \emph {et~al.}(2017)\citenamefont
  {Williams}, \citenamefont {Bodmann},\ and\ \citenamefont
  {Kouri}}]{williams2017fourier}%
  \BibitemOpen
  \bibfield  {author} {\bibinfo {author} {\bibfnamefont {C.~L.}\ \bibnamefont
  {Williams}}, \bibinfo {author} {\bibfnamefont {B.~G.}\ \bibnamefont
  {Bodmann}}, \ and\ \bibinfo {author} {\bibfnamefont {D.~J.}\ \bibnamefont
  {Kouri}},\ }\href@noop {} {\bibfield  {journal} {\bibinfo  {journal} {Journal
  of Fourier Analysis and Applications}\ }\textbf {\bibinfo {volume} {23}},\
  \bibinfo {pages} {660} (\bibinfo {year} {2017})}\BibitemShut {NoStop}%
\bibitem [{\citenamefont {Williams}(2017)}]{williams2017}%
  \BibitemOpen
  \bibfield  {author} {\bibinfo {author} {\bibfnamefont {C.~L.}\ \bibnamefont
  {Williams}},\ }\emph {\bibinfo {title} {From Generalized Fourier Transforms
  to Coupled Supersymmetry}},\ \href@noop {} {Ph.D. thesis},\ \bibinfo
  {school} {University of Houston} (\bibinfo {year} {2017})\BibitemShut
  {NoStop}%
\bibitem [{\citenamefont {Kouri}\ \emph {et~al.}(2018)\citenamefont {Kouri},
  \citenamefont {Pandya}, \citenamefont {Williams}, \citenamefont {Bodmann},\
  and\ \citenamefont {Yao}}]{kouri2018point}%
  \BibitemOpen
  \bibfield  {author} {\bibinfo {author} {\bibfnamefont {D.}~\bibnamefont
  {Kouri}}, \bibinfo {author} {\bibfnamefont {N.}~\bibnamefont {Pandya}},
  \bibinfo {author} {\bibfnamefont {C.~L.}\ \bibnamefont {Williams}}, \bibinfo
  {author} {\bibfnamefont {B.~G.}\ \bibnamefont {Bodmann}}, \ and\ \bibinfo
  {author} {\bibfnamefont {J.}~\bibnamefont {Yao}},\ }\href@noop {} {\bibfield
  {journal} {\bibinfo  {journal} {Applied Mathematics}\ }\textbf {\bibinfo
  {volume} {9}},\ \bibinfo {pages} {178} (\bibinfo {year} {2018})}\BibitemShut
  {NoStop}%
\bibitem [{\citenamefont {{Kouri}}\ \emph {et~al.}(2017)\citenamefont
  {{Kouri}}, \citenamefont {{Zhang}},\ and\ \citenamefont {{Zhang}}}]{8270495}%
  \BibitemOpen
  \bibfield  {author} {\bibinfo {author} {\bibfnamefont {D.~J.}\ \bibnamefont
  {{Kouri}}}, \bibinfo {author} {\bibfnamefont {M.~M.}\ \bibnamefont
  {{Zhang}}}, \ and\ \bibinfo {author} {\bibfnamefont {D.~S.}\ \bibnamefont
  {{Zhang}}},\ }in\ \href {\doibase 10.1109/ICSPCS.2017.8270495} {\emph
  {\bibinfo {booktitle} {2017 11th International Conference on Signal
  Processing and Communication Systems (ICSPCS)}}}\ (\bibinfo {year} {2017})\
  pp.\ \bibinfo {pages} {1--5}\BibitemShut {NoStop}%
\bibitem [{\citenamefont {Williams}\ \emph {et~al.}(2014)\citenamefont
  {Williams}, \citenamefont {Bodmann},\ and\ \citenamefont
  {Kouri}}]{williams2014fourier}%
  \BibitemOpen
  \bibfield  {author} {\bibinfo {author} {\bibfnamefont {C.~L.}\ \bibnamefont
  {Williams}}, \bibinfo {author} {\bibfnamefont {B.~G.}\ \bibnamefont
  {Bodmann}}, \ and\ \bibinfo {author} {\bibfnamefont {D.~J.}\ \bibnamefont
  {Kouri}},\ }\href@noop {} {\bibfield  {journal} {\bibinfo  {journal} {arXiv
  preprint arXiv:1403.4168}\ } (\bibinfo {year} {2014})}\BibitemShut {NoStop}%
\bibitem [{\citenamefont {Haberman}(1983)}]{haberman1983elementary}%
  \BibitemOpen
  \bibfield  {author} {\bibinfo {author} {\bibfnamefont {R.}~\bibnamefont
  {Haberman}},\ }\href@noop {} {\emph {\bibinfo {title} {Elementary applied
  partial differential equations}}},\ Vol.\ \bibinfo {volume} {987}\ (\bibinfo
  {publisher} {Prentice Hall Englewood Cliffs, NJ},\ \bibinfo {year}
  {1983})\BibitemShut {NoStop}%
\bibitem [{\citenamefont {Li}(2016)}]{li2016powers}%
  \BibitemOpen
  \bibfield  {author} {\bibinfo {author} {\bibfnamefont {C.}~\bibnamefont
  {Li}},\ }\href@noop {} {\bibfield  {journal} {\bibinfo  {journal} {J. Fract.
  Calc. Appl}\ }\textbf {\bibinfo {volume} {7}},\ \bibinfo {pages} {12}
  (\bibinfo {year} {2016})}\BibitemShut {NoStop}%
\bibitem [{\citenamefont {Jarad}\ \emph {et~al.}(2018)\citenamefont {Jarad},
  \citenamefont {Adjabi}, \citenamefont {Baleanu},\ and\ \citenamefont
  {Abdeljawad}}]{jarad2018defining}%
  \BibitemOpen
  \bibfield  {author} {\bibinfo {author} {\bibfnamefont {F.}~\bibnamefont
  {Jarad}}, \bibinfo {author} {\bibfnamefont {Y.}~\bibnamefont {Adjabi}},
  \bibinfo {author} {\bibfnamefont {D.}~\bibnamefont {Baleanu}}, \ and\
  \bibinfo {author} {\bibfnamefont {T.}~\bibnamefont {Abdeljawad}},\
  }\href@noop {} {\bibfield  {journal} {\bibinfo  {journal} {Advances in
  Difference Equations}\ }\textbf {\bibinfo {volume} {2018}},\ \bibinfo {pages}
  {407} (\bibinfo {year} {2018})}\BibitemShut {NoStop}%
\bibitem [{\citenamefont {Anteneodo}\ \emph {et~al.}(2006)\citenamefont
  {Anteneodo}, \citenamefont {Dias},\ and\ \citenamefont
  {Mendes}}]{anteneodo2006long}%
  \BibitemOpen
  \bibfield  {author} {\bibinfo {author} {\bibfnamefont {C.}~\bibnamefont
  {Anteneodo}}, \bibinfo {author} {\bibfnamefont {J.}~\bibnamefont {Dias}}, \
  and\ \bibinfo {author} {\bibfnamefont {R.}~\bibnamefont {Mendes}},\
  }\href@noop {} {\bibfield  {journal} {\bibinfo  {journal} {Physical Review
  E}\ }\textbf {\bibinfo {volume} {73}},\ \bibinfo {pages} {051105} (\bibinfo
  {year} {2006})}\BibitemShut {NoStop}%
\bibitem [{\citenamefont {Toscani}(2005)}]{toscani2005central}%
  \BibitemOpen
  \bibfield  {author} {\bibinfo {author} {\bibfnamefont {G.}~\bibnamefont
  {Toscani}},\ }\href@noop {} {\bibfield  {journal} {\bibinfo  {journal}
  {Journal of Evolution Equations}\ }\textbf {\bibinfo {volume} {5}},\ \bibinfo
  {pages} {185} (\bibinfo {year} {2005})}\BibitemShut {NoStop}%
\bibitem [{\citenamefont {Schw{\"a}mmle}\ \emph {et~al.}(2008)\citenamefont
  {Schw{\"a}mmle}, \citenamefont {Nobre},\ and\ \citenamefont
  {Tsallis}}]{schwammle2008q}%
  \BibitemOpen
  \bibfield  {author} {\bibinfo {author} {\bibfnamefont {V.}~\bibnamefont
  {Schw{\"a}mmle}}, \bibinfo {author} {\bibfnamefont {F.~D.}\ \bibnamefont
  {Nobre}}, \ and\ \bibinfo {author} {\bibfnamefont {C.}~\bibnamefont
  {Tsallis}},\ }\href@noop {} {\bibfield  {journal} {\bibinfo  {journal} {The
  European Physical Journal B}\ }\textbf {\bibinfo {volume} {66}},\ \bibinfo
  {pages} {537} (\bibinfo {year} {2008})}\BibitemShut {NoStop}%
\bibitem [{\citenamefont {Kouri}\ \emph {et~al.}(2017)\citenamefont {Kouri},
  \citenamefont {Pandya}, \citenamefont {Williams}, \citenamefont {Bodmann},\
  and\ \citenamefont {Yao}}]{kouri2017point}%
  \BibitemOpen
  \bibfield  {author} {\bibinfo {author} {\bibfnamefont {D.~J.}\ \bibnamefont
  {Kouri}}, \bibinfo {author} {\bibfnamefont {N.~N.}\ \bibnamefont {Pandya}},
  \bibinfo {author} {\bibfnamefont {C.~L.}\ \bibnamefont {Williams}}, \bibinfo
  {author} {\bibfnamefont {B.~G.}\ \bibnamefont {Bodmann}}, \ and\ \bibinfo
  {author} {\bibfnamefont {J.}~\bibnamefont {Yao}},\ }\href@noop {} {\bibfield
  {journal} {\bibinfo  {journal} {arXiv preprint arXiv:1708.00074}\ } (\bibinfo
  {year} {2017})}\BibitemShut {NoStop}%
\end{thebibliography}%

\end{document}